# Business Process Model for Interoperability Improvement in the Agricultural Domain Using Digital Twins

*Completed Research Paper*

**Emily Calvet**
Fraunhofer Institute for Experimental
Software Engineering IESE
Kaiserslautern, Germany
emily.calvet@iese.fraunhofer.de

**Rodrigo Falcão**
Fraunhofer Institute for Experimental
Software Engineering IESE
Kaiserslautern, Germany
rodrigo.falcao@iese.fraunhofer.de

**Lucineia Heloisa Thom**
Informatics, Federal University of Rio Grande do Sul
Porto Alegre, Rio Grande do Sul, Brazil
lucineia@inf.ufrgs.br

## Abstract

*A farm generates a lot of data from various systems, which is then stored in a distributed manner, usually in non-standardized formats, which bears the risk of data inconsistencies. This work addresses this issue by using business process management (BPM) to demonstrate that the use of digital twins (DTs) can improve interoperability between services in the agriculture domain. Steps from the BPM lifecycle were applied to a farming use case in Germany. First, the as-is business process model was discovered and modeled without DTs, analyzed and then redesigned into the to-be model according to the DT integration. The to-be model showed a reduction in the number of tasks needed to be performed by the farmer as well as an improvement of process data quality, interoperability, and efficiency. Finally, a comparison of the' average processing times of both models with the help of process simulation revealed improvements in the to-be process.*

## Introduction

Agriculture is one of the most important practices for human survival and has undergone various revolutions and adaptations in order to better supply all emerging demands. Among the challenges posed to agriculture over time are the demand for high-quality groceries, traceable food, and more sustainability (Doerr and Nachtmann 2022). Nowadays, farms in Europe have sensors all over, gathering data on weather, soil composition, plant conditions, machinery, and so on (EIP-AGRI 2015). All this data is analyzed and stored in agricultural domain-specific systems, usually Farm Management Information Systems (FMIS), that help the farmer to understand what the data means and that provide decision support for the next steps to take. An FMIS can be seen as an Enterprise Resource Planning (ERP) software for agricultural activities, as it is designed to store data and manage a farm's resources, e.g., fields, machinery, and employees (Doerr and Nachtmann 2022).

Due to the lack of standardization of agricultural data formats in the early years, many companies developed their own agriculture-specific software using the data formats they considered best, which has resulted in a cluster of systems that are not interoperable with each other (IESE 2020). All these available systems and their data compose the current agricultural data space (ADS). Farms have diverse processes that are tackled by specific systems, meaning that a farmer often needs to deal with more than one system to manage their farm as a whole. Furthermore, the farm data becomes distributed since these applications are not encouraged to share data with each other, as their providers seek to offer completely proprietary systems (EIP-AGRI 2015). Consequently, this leads to future data inconsistency problems and to more effort for the farmer to keep their data in sync.

A way to solve these issues would be to make FMISs more collaborative, interoperable and scalable with each other by adopting a data space approach that integrates different FMISs and makes agricultural data more shareable (Zaninelli and Pace 2018). This can be achieved through standardization of farm workflows and farm data achieved by the collaboration of technical stakeholders and international organizations, as done in other domains such as the medical domain (Zaninelli and Pace 2018).

In the research project COGNAC[1], these challenges are addressed through the use of digital twins (DTs) to improve interoperability and data security, taking as a reference their use in Industry 4.0 (IESE 2021). The purpose of the project is to build a greatly interoperable and secure ADS. This ADS is enabled by a digital platform, the ADS Platform, to provide an interface between the participants of the ecosystem and implement their interactions according to set rules (IESE 2020). A DT is a digital version of an asset with a two-way connection to its physical twin (QI et al. 2021). Initially, the project is studying the creation of digital field twins, as fields play a major role in crop farming. The field twin should encapsulate all data belonging to a field that has great relevance, making it easy for services to find all necessary data for their execution in one place.

The purpose of this work is to evaluate and demonstrate the potential improvement of interoperability among existing systems, which can be achieved by integrating the ADS Platform and the DTs. This is done in the context of a sugar beet farming use case studied by COGNAC. Our research question is formulated as follows:

*RQ: Can the use of a digital field twin improve interoperability among existing systems in the agriculture domain?*

To address this question, we applied the Business Process Management (BPM) lifecycle steps to discover, analyze, and redesign the sugar beet farming process. BPM is an already consolidated and largely adopted discipline that brings detailed insights into the process landscape and uncovers optimization potential (Dumas et al. 2018). We modeled the process as it is currently done (as-is) in Germany and then analyzed it to find its improvement points. Based on the findings, we modeled the process as it should be executed (to-be) to obtain the improvements following the integration of the DTs. The modeling for both steps was done with the Business Process Model and Notation (BPMN) (OMG 2021). The findings show that an improvement in data quality and interoperability could be achieved by centralizing the data in the DTs. We performed further evaluations using the simulation tool from Signavio[2] to compare the processing time of the processes. The results show that the redesigned model had improved efficiency compared to the as-is process.

This paper is organized as follows. First, it covers the theoretical fundamentals discussed and implemented in this work by defining the concepts used and reporting on related work. Next, the methodology we used is explained, and its usage is described. Finally, the results are discussed, and conclusions and comments on future work are given.

## Background and Related Work

BPM is a discipline that involves methods, principles, and tools for not only discovering the processes within an organization but for documenting, analyzing, monitoring, and improving them in order to gain the best value for the target stakeholders (Dumas et al. 2018). A process is a chronological sequence

---

[1] https://cognitive-agriculture.de
[2] https://www.signavio.com/

composed of a number of events and activities performed by actors. This discipline encloses the entire business process lifecycle. BPM is an already consolidated and widely practiced discipline that aims at reducing costs, execution time, and errors, and increasing control over process execution, thus also improving process quality (Dumas et al. 2018). BPM has a lifecycle that structures the concepts, methods, techniques, and tools of the BPM discipline and provides a sequential order of steps to be performed to manage the process and add value to it (Dumas et al. 2018). The BPM lifecycle comprises six phases:

*Process identification* is the initial phase where a given problem is addressed by recognizing the processes relevant to it, followed by the identification of the relevant processes that should be taken further into the subsequent phases.

*Process discovery* aims at understanding and modeling the identified processes in detail. The main method categories for information gathering are *evidence-based discovery* and *interview-based discovery*. The former collects process information according to documentation, on-site observation or event logs from automated process execution. The latter depends on the collaboration of process participants and domain experts during an interview-modeling validation cycle. It is a method that takes time and effort, but provides valuable insight into how the process is seen and done by its participants. This is the method utilized in this work. Finally, a process model is generated in the form of a diagram that represents the discovered process, which is called *as-is*.

*Process analysis* inspects the model in search of vulnerabilities and points of improvements that are then documented in an *issue register*. It has two categories of techniques, *quantitative* and *qualitative*. In quantitative techniques, previously chosen process performance measures can be used to analyze the as-is process and identify its issues. The issues can then be documented, prioritized, and investigated in order to arrive at a solution. For example, if the most important measure is time, the analyst needs to gather information on how much time each activity takes and assert whether it is damaging or not to the value given by the outcome of the process. Qualitative approaches are aimed at finding issues that impact the process in a manner that is difficult to quantify, such as customer satisfaction or data quality. One of the most commonly used qualitative techniques is *Waste Analysis* (Dumas et al. 2018). It focuses on finding and reducing activities that are dispensable from the process based on the value it adds to the client of the process. There are seven waste categories that Dumas et al. (2018) grouped into three:

- Move: *transportation* and *motion*, waste related to movement;
- Hold: *inventory* and *waiting*, waste caused by holding something;
- Overdo: *defects*, *overprocessing*, and *overproduction*, waste created by doing more than necessary to deliver the value to the client of the process.

*Process redesign* uses the documented issues found in the previous phase and focuses on identifying and evaluating possible solutions based on the chosen methodology. There is a whole spectrum of methods to be picked from, called the Redesign Orbit, depending on the individual case of each organization (Dumas et al. 2018). The methods can be either *transactional* or *transformational*, *creative* or *analytical*, *inward-looking* or *outward-looking*. This means that this orbit has three dimensions and a method can belong to one of the two types of each dimension. Transactional methods take the current process as an initial point of examination and aim at gradual overall improvement. Transformational methods, on the other hand, seek revolutionary innovation for the organization, usually by building an entirely new process. The difference between creative and analytical methods is that, while the former utilizes mostly human creativity with group activities such as workshops, the latter has a mathematical basis that uses quantitative techniques to analyze the process. Finally, an inward-looking method takes the viewpoint of the organization and an outward-looking method has an outsider's perspective, e.g., a customer or a third-party organization. The result of this stage is the remodeled process, called *to-be*, with the best solutions selected for the organization's problem.

The purpose of *process implementation* is to actually implement the organizational and technical changes needed to execute the to-be process.

In the *process monitoring* phase, degradation of the process is avoided by collecting and evaluating relevant data according to the expected performance quality set to ascertain whether the process still provides the best value to the organization.

*Business Process Model and Notation*

BPMN is an ISO standard released by the Object Management Group (OMG) and has become the de-facto notation for business process diagrams, providing a simple means of communicating process information to its various stakeholders (OMG 2022). It is a flow chart-based notation created in agreement with several modeling tool vendors to use a single notation able to support complex scenarios and to benefit both technicians and business users (Celestrini et al. 2019). Furthermore, modeling the process in a notation, provides the possibility to document, automate, simulate, and verify the process in a Business Process Management System (BPMS). We will describe its elements below, with a focus on the elements used in this work based on the BPMN 2.0 specification (OMG 2021).

*Flow Objects* are the main elements used to define the behavior of the business process. They are composed of *Events*, *Activities,* and *Gateways*. An *Event* is an atomic (without time duration) situation that happens during a process and can trigger or result in something. There are three sub-categories of events: one that starts a process (*Start*), one that affects the flow of a process (*Intermediate*), and one that ends a process (*End*). An event can have a type to better express its context. This work uses the *Message* type. A message trigger event implies that a message was received from a process participant and triggers the start of a process (start event) or of an activity (intermediate event).

*Activities* are work that takes time. A *Task* is an atomic activity. Markers can also be added to a task to facilitate the understanding of the task type. In this work, the task types used are *User Tasks* and *Manual Tasks*. Both of them are performed by a human, but a user task is an activity executed with the assistance of a software application through a UI, while a manual task is done entirely by the person performing it.

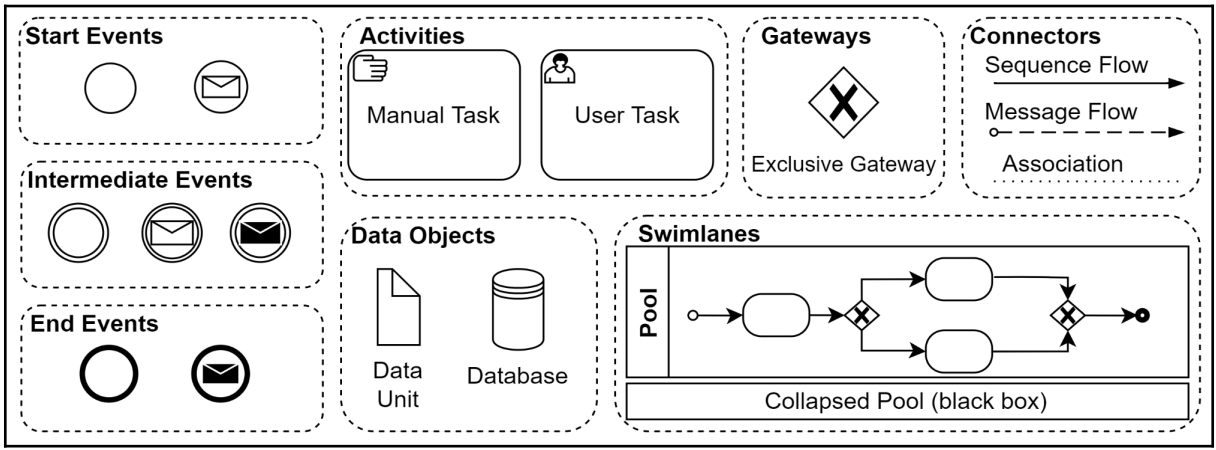

**Figure 1: BPMN elements.**

*Gateways* are used to control how the process diverges and converges during its flow. In this work, the *Exclusive* type is used, which represents a decision on the process path where only one way can be chosen for the current process instance in execution based on the decision made at the gateway.

*Data* is represented by data objects that demonstrate how information is used or produced within a process. They can define inputs and outputs while also having a state to present how they may have changed. In this work, the data objects represent digital information related to a field relevant to a farm's processes. A data store is a container for data objects that need to be persisted for longer than a process duration, e.g., a database.

*Connecting Objects* connect the flow objects to each other or to information by linking them with arrows or lines. They are divided into three sub-categories. The first is *Sequence Flow*, which depicts the order of the activities. The second is *Message Flow*, which demonstrates the messages exchanged between participants of a process. The third is *Association*, which represents the link between information and artifacts on the one side and other elements on the other side. The association can be directed when appropriate.

*Swimlanes* were created to help distinguish roles and their interactions within a process. They can be *Pools* or *Lanes*. Pools represent participants, as a role (e.g., "customer") or as a business entity (e.g., "OMG"). They can simply be a black box or contain a process. The interaction between pools is taken care of solely by the message flow. A lane is a sub-partition in a process, occasionally within a pool, representing an organizational role (e.g., "manager").

### *Process Model Quality*

There are three main quality aspects: *syntactic, semantic, and pragmatic quality*. The assurance procedure for each of these is called verification, validation, and certification, respectively (Dumas et al. 2018). Syntactic quality refers to how the model complies with the rules of the modeling language used. Therefore, it is about structural correctness. Semantic quality is more complicated to check, since it is related to the compatibility of the model with its real-world process and has no specific rules to be followed. Thus, it should be checked by talking it through with one of the participants of the process or a domain expert. *Pragmatic quality* refers to how the model can be useful for its purposes. This work follows the Seven Process Modeling Guidelines (7PMG) defined by Mendling, Reijers, and Aalst (2010), as well as the guidelines defined by Avila and Thom (2019). Both of them define rules to prevent the model from getting too large or too confused to understand, and propose naming conventions for the elements to avoid language ambiguities.

### *Digital Twin*

With the advances of technology in areas such as the Internet of Things, Big Data Analytics, Artificial Intelligence, and Cloud Computing, the convergence of the physical and virtual worlds becomes increasingly palpable (QI et al. 2021). The term Digital Twin (DT) was first introduced by Grieves in 2003 at a time when those technologies were still in development (Grieves 2014).

Essentially, a DT determines the existence of a physical entity and its complete digital representation in total communication harmony, sharing data through bidirectional interactions (QI et al. 2021). This means that everything that happens in one twin has to be replicated in near real-time in the other twin. It is an ongoing cycle of monitoring the status and analyzing it to recognize complexities, detecting irregular behaviors, simulating changes, predicting events, and so on, towards the collective evolution of the twins. For example, previously simulated tasks can be triggered in the virtual twin and then performed on its physical twin. There is a lot of data and knowledge from the physical world of the real entity involved to ensure that the virtual entity is as true as it can get to its physical twin. By analyzing the data history, e.g., past problems, simulations, and forecasts, DTs can greatly improve decision-making, productivity, and efficiency in various activities. Furthermore, they can reduce the cost and time spent for on-site evaluations and assessments (QI et al. 2021). DTs not only create an integrated and centralized knowledge base, but also combine this information with meta-information, allowing a semantic description of an asset (Haße et. al 2020). For this reason, it is a valid approach for solving the problem of data disruption between distributed systems (Haße et. al 2020).

Although DTs are still in the modeling stage in the agriculture domain (Pylianidis, Osinga, Athanasiadis 2021), they are already widely used in the areas of production, prognostics, and health management (PHM), providing more reliability, flexibility, and predictability to their processes (Tao et al. 2019). The main challenges for DT implementation in the agricultural domain include: dealing with complex living subjects like animals and plants, slow adoption of technology, and lack of financial incentives (Pylianidis, Osinga, Athanasiadis 2021). Other challenges include the successful implementation of the real-time two-way connection between the twins and the handling of large amounts of data, data variety, connection latency, and online processing of data (Rasheed, San, Kvamsdal 2020).

### *Related Work*

The application of DTs in agriculture is still relatively new. Most were developed only to a conceptual level (Pylianidis, Osinga, Athanasiadis 2021). The research methodology used in this work consisted of finding publications in the agriculture domain that comprise DTs or BPM or both, if available. Furthermore, we reviewed the identified publications to see if they also discussed interoperability in any way and integration possibilities with other services. The related work found is described below, followed by Table 1, which summarizes which work has which characteristic.

Pylianidis, Osinga, and Athanasiadis (2021) did a literature review of DTs in agriculture and their benefits to estimate to what extent they have been applied. The results show that there are already some applications of DTs in agriculture, but they are still in early stages and not thoroughly designed or deployed as in other domains. However, the authors provide a potential roadmap of applications to facilitate their adoption similar to the DTs currently used in other domains.

Rupnik et al. (2021) tackles the interoperability issue by creating a reference standard process model for agriculture (RSPMA) in order to help software companies, which commonly do not have enough knowledge of agricultural processes, develop their farming-specific systems. The authors acknowledge the hazards that the lack of business process and data standardization can cause to costs, development, and competitiveness. Moreover, they state that a reference standard process model can improve the efficiency, effectiveness, and interoperability of the systems since the model is already validated and avoids common faults. Their first version was validated by an international panel of twenty experts in the agriculture domain. This work does not use the DT concept.

Cestari et. al (2020) proposes an integration Platform as a Service (iPaaS) architecture to improve the integration, interoperability, extensibility, and automated decision-making capabilities in the agriculture domain. The DT concept was not used in this work. They tested their work with a case study on the grain storage process, comparing it with the current process as performed by humans. The objective was to reduce energy consumption and the need for human operators. Automated decision-making was achieved through BPM engines and machine learning modules.

Zaninelli and Pace (2018) modeled the milking processes of a small farm to organize its activities as workflows and drive the software development of a web-based FMIS for the traceability of milk, called O3-Farm. BPMN was employed for the modeling of the O3-Farm processes. For the validation, they had five professional farmers test the old version and the O3-Farm version of a management system according to basic tasks and then asked them to provide their opinion in a questionnaire. The results showed that O3-Farm provided all the data needed to trace milk yields as well as new features that improved usability, portability, and efficiency. The new application also offers the possibility of integration with other services due to its better shareability of agricultural data. This research, which is the oldest we found, mentions a farm software environment (FSE), which proposes the concept of making FMISs more interoperable, scalable, linkable, and collaborative. The authors mention that this model would be possible by integrating software applications and shareable agricultural data, something that could be achieved by standardization of farm workflows with BPM, standardization of farm data, and the collaboration of key stakeholders as described by the authors. This concept is very similar to the ADS from COGNAC.

Keates (2019) examined the installment of BPM in Internet of Things (IoT) implementations in cattle farming in Australia with the intent to achieve more value by improving decision support. They utilized the BPM lifecycle defined by Dumas et al. (2018) and BPMN to model the processes specifically during the process redesign phase so that they could be executed later in a BPMS (Dumas et al. 2018). The authors concluded that it is very difficult to gather data from different systems and IoT solutions (lack of interoperability) and that this should be considered in the near future by these solutions providers, for the benefit of the customer. Furthermore, they used the concept of DTs to virtualize the cattle and be able to simulate data missing from the sensors' outputs. However, they do not show any intention to integrate this with other solutions and use DTs to increase interoperability itself.

Celestrini et al. (2019) proposed an IoT solution integrated with BPMN to manage the production process in Controlled Environment Agriculture. They developed an architecture using business process rules for managing IoT devices through BPMN. The work was validated by applying it to a case study to assist in vegetable production, with business rules modeled by a domain expert. The authors concluded that BPMN improved process modeling, thus providing more control of the production process and enhancing the decision-making of the producers. Moreover, BPMN enabled their architecture to become more flexible in the event of crop changes in the same environment without the need to change the code. They overcame the interoperability issue between IoT sensors through a component responsible for converting all raw data received, but they did not utilize the DT concept or intend to integrate their solution with other services.

Alves et al. (2019) presents an initial development of a DT for a smart farm by creating a cyber-physical system (CPS) that gathers data from the soil and displays it on a dashboard. They integrated IoT to control an irrigation system based on the farmer's or the AI's decision. The work needs to be further developed to integrate multiple systems in order to represent all the processes involved on the farm. However, the authors do not state any intent to integrate their work with third-party systems or make it interoperable.

| Research/ Characteristic | Digital Twin | Interoperability | Intent to integrate with other services |
|---|---|---|---|
| Rupnik et al. (2021) | | X | X |
| Cestari, Ducos, and Exposito (2020) | | X | |
| Zaninelli and Pace (2018) | | X | X |
| Keates (2019) | X | X | |
| Celestrini et al. (2019) | | X | |
| Alves et al. (2019) | X | | |

**Table 1. Relevant characteristics of related work**

Given the related work research, we see that only few authors use the DT concept or have the intent to improve interoperability, standardize farm data, and integrate their work into the current agricultural data space to leverage it and collaborate with existing systems and services. Therefore, there is a need to study solutions to bring the agriculture domain to a more standardized and collaborative state between its services, as already done in other domains. Among other improvements, the COGNAC project aims at achieving more interoperability through data standardization and integration between existing systems in agriculture through the use of digital field twins. The objective of this work is to demonstrate the improvement of interoperability and the integration that is possible with the COGNAC solution by applying it to the already consolidated discipline of BPM.

## Research Method

In order to address our research question and evaluate the interoperability improvement brought about by the integration of DTs into the ADS proposed by COGNAC, we studied the use case of how the sugar beet farming process is currently done in Germany. To always ensure the best product value, it is no longer sufficient to evaluate only the final product, but also to evaluate its production process (Rupnik et al. 2021). With the BPM discipline, this process can be discovered, analyzed, redesigned, and monitored in order to follow market, legal and environmental changes, carrying out process reengineering as needed to keep the best product value desired by its stakeholders. Furthermore, modeling and documenting the process facilitates its digitalization, standardization, and improvement because it allows separating the logical layer from the technical implementation (Rupnik et al. 2021). Therefore, the BPM initiative will be used to analyze and improve the sugar beet farming process through the execution of the BPM lifecycle steps and, within each step, the appropriate method in relation to the use case.

In this work, we executed three consecutive steps of the BPM lifecycle that followed the process identification in a sugar beet farming use case and then evaluated its improvement. The identification step had already been done in the project, and we did not execute the implementation and monitoring phases, as this work aims primarily at proofing this concept before its implementation. The methodology is shown in Figure 2. First, the as-is process model of the whole process of sugar beet farming was discovered (Process Discovery) through a series of interviews with the designated agriculture domain expert from COGNAC. The interviews were semi-structured, with prepared questions to be answered about the sugar beet farming process activities. Then the process was modeled using BPMN and following the seven process modeling guidelines (7PMG) that assist in pragmatic quality assurance. The model was syntactically verified by the BPMN modeling tool Signavio, which was evaluated by Dias et al. (2019) to be the tool that provides the best support in process verifications. Additionally, the model was structurally verified by two BPM experts from the UFRGS BPM research group. Derived from the model, a complete textual description of the process was also created to support the expert's understanding during the semantic validation step. This operation was performed iteratively until the model was considered semantically correct by the expert.

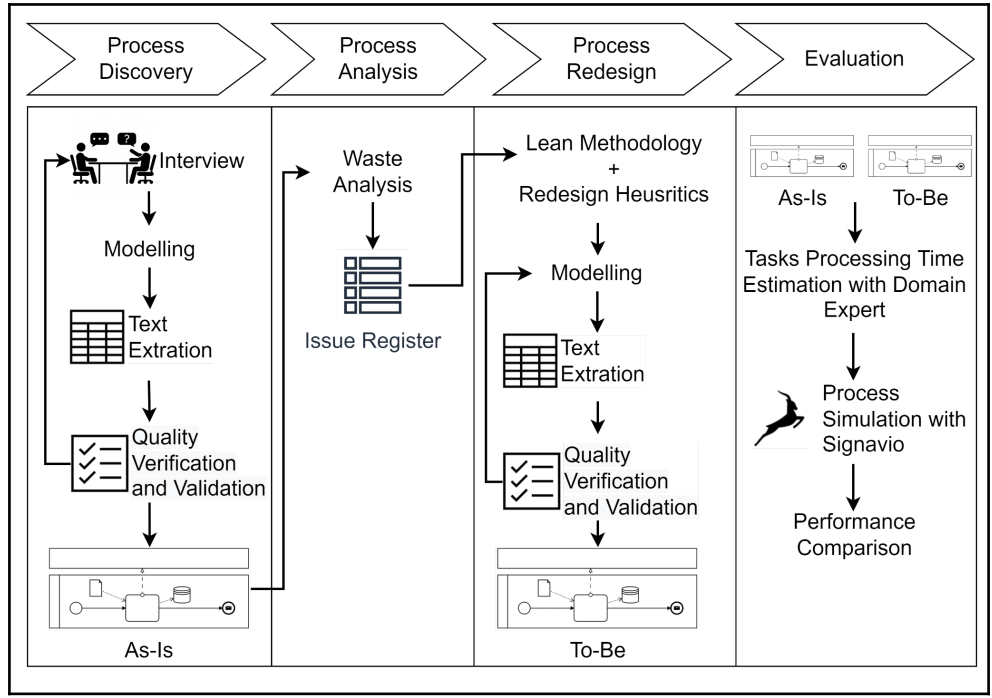

**Figure 2: Methodology view.**

To create precise textual process descriptions from the designed model, a sentence template was used to translate the control flow information into text. The sentence template used was adapted by **Silva et al. (2019)** and is shown in Table 1. With this method, the cooperation between the process analyst and domain experts is facilitated as the process description is constructed using a controlled structure based on the features of the process modeling notation (Silva et. al 2019). Moreover, the ambiguities of natural language are minimized for the sake of creating a better understanding between process model and description. Using this approach, the model was completely paraphrased in textual form while maintaining its semantic meaning. This enabled the process analyst to validate it with the domain experts, with fewer misunderstandings caused by the lack of knowledge of the modeling language on their part. This approach was applied manually for both the as-is and the to-be process.

| Element | Type | Sentence Templates |
|---|---|---|
| Start | Sequence | The <*process*> process starts when… |
| Sequence | Atomic 1<br>Atomic 2<br>Atomic 3 | Then, …<br>After < that \| *illness* manifests>, …<br>Subsequently, … |
| End | Sequence | The process ends with … |
| Exclusive Choice | Split 1<br>Split 2<br>Join | The < *cond.* > may either be < *first* >, or < *second* >, …<br>The < *role* > may either be < *first* >, or < not \| *second* >. …<br>In any < case \| of these cases >, … |
| Inclusive Choice | Split<br>Join | …, <*number*> alternative procedures may be executed.<br>Afterwards, … |

*Business Process Model for Interoperability Improvement in Agriculture*| Choice Paths | Ordinal<br>Conditional | In the < *first* \| *second* \| *third*... >procedure, . . .<br>If < *cond.* >, ... |
|---|---|---|
| Parallelism | Split<br>Join<br>Path 1<br>Path 2 | ..., < *number* > procedures are executed in an arbitrary order.<br>After each case, ...<br>In the meantime, ...<br>At the same time, ... |
| Loop | Join 1<br>Join 2 | If required, < *role* > repeats the latter steps and continues with ...<br>Once the loop is finished, ... |

**Table 2. Sentence templates. Source: Silva et al. (2019)**

After approval by the domain expert, the model can be analyzed for vulnerabilities and points of improvement to be documented in an issue register (Process Analysis). Based on the results found, the process redesign step remodels the process following the Lean methodology and suitable redesign heuristics to improve or eradicate the found problems. The produced to-be model then goes through the same verification and validation operation as the as-is model, only this time it is semantically validated by an expert in the project through its model and text description until a satisfactory state is reached.

Finally, both the as-is and the to-be models are simulated in the Signavio tool using processing times estimated by the domain expert for each activity. The average processing times of the processes are then compared to verify if there is also an improvement in time.

The following sections describe how each step of the method was executed and its results.

## Process Discovery

In this step, an interview-based approach was taken with the designated agriculture domain expert from COGNAC. We performed a semi-structured interview where the expert was asked prepared questions for each activity of the process. In advance, we already had a sketch of the process based on the standard steps for planting in crop farming, which comprise: tillage, seeding, weed control, fertilization, plant protection, and harvesting. Then the prepared questions were asked in each interview and new activities or flows were added to the process as the expert answered and introduced new and relevant information about the sugar beet farming process currently performed in Germany. Moreover, the expert answered clarifying questions when needed by the interviewer. One interviewee may seem insufficient to provide data triangulation; however; the expert in question is a specialist in the agricultural domain and is designated to seek and provide accurate information on the project. The prepared questions asked were:

- Q1: Which information is required for this activity?
- Q2: Which information is produced by this activity?
- Q3: How is this activity performed?
- Q4: Is this activity performed with the support of a system?
- Q5: Which kind of FMIS is needed for this activity?
- Q6: Where is the information required by this activity stored?
- Q7: Where is the information produced by this activity stored?

The process was modeled according to the expert as it is currently executed for farmers who utilize digital tools. After each interview, the process model was updated, verified, and then sent back to the expert along with its paraphrased textual description for semantic validation. In other words, if the text was considered correct by the expert, the model was also correct; therefore, the model was considered semantically correct. Pragmatic quality was assured by following the 7PMG. Moreover, the syntactic quality of the model was verified by two BPM experts and the Signavio tool. The result is a model of an example farm that has three different systems to conduct its operations and that represents how it would handle the sugar beet farming process.

*Pacific Asia Conference on Information Systems 2022*

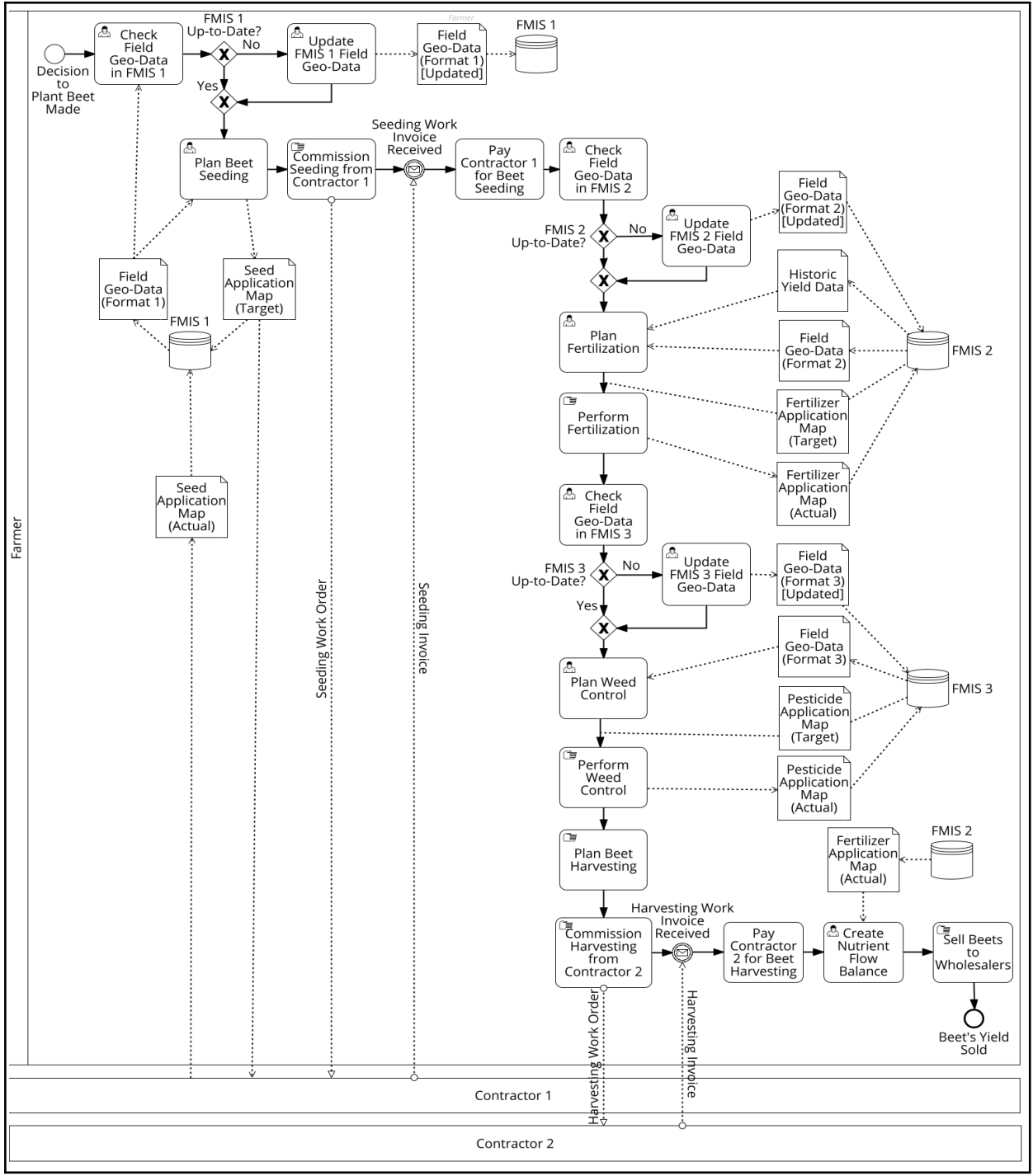

**Figure 3. Sugar beet farming as-is process model.**

The extracted textual description and the final as-is model version, is presented in the following. "The *Sugar Beet Farming As-Is* process starts when the farmer decides to plant sugar beet in the field (*Start Event: Decision to Plant Beet Made*). Next, the farmer checks if the *Field Geo-Data* is up-to-date in *FMIS 1* (*Task: Check Field Geo-Data in FMIS 1*). The system *Field Geo-Data* can either be up-to-date or out-of-date depending on the field geometry desired for the current sugar beet farming. If the *FMIS 1 Field Geo-Data* is out-of-date, the farmer then updates it in *FMIS 1*, which stores the *Field Geo-Data (Format 1) [Updated]* in its database. After that, the farmer can move on to the planning of sugar beet

*Business Process Model for Interoperability Improvement in Agriculture*seeding (*Task: Update FMIS 1 Field Geo-Data*). If the *FMIS 1 Field Geo-Data* is up-to-date, the farmer can go directly to the planning of sugar beet seeding.

Subsequently, the farmer plans the seeding operation with the support of an FMIS with seeding features (*FMIS 1*) that contains their *Field Geo-Data* in the format required by it *(Field Geo-Data (Format 1))* in its database. The activity results in a target application map (*Seed Application Map (Target)*) for the sugar beet seeds that is also stored in *FMIS 1's* database (*Task: Plan Beet Seeding*). After that, the farmer contracts a contractor (*Contractor 1*) of their choice to perform the seeding, who notifies the farmer when the job is done (*Task: Contract Seeding from Contractor 1*). When the farmer receives the notification with the invoice for the completed work and the *Seed Application Map (Actual)* (*Intermediate Catching Message Event: Seeding Work Invoice Received*), they pay *Contractor 1* for the work done (*Task: Pay Contractor 1 for Beet Seeding*).

Afterwards, the farmer checks if the *Field Geo-Data* is up-to-date in *FMIS 2* (*Task: Check Field Geo-Data in FMIS 2*). If the *FMIS 2 Field Geo-Data* is out-of-date, the farmer then updates it in *FMIS 2* which stores the *Field Geo-Data (Format 1) [Updated]* in its database. After this, they can move on to the planning of sugar beet fertilization *(Task: Update FMIS 2 Field Geo-Data)*. If the *FMIS 2 Field Geo-Data* is up-to-date, the farmer can go directly to the planning of sugar beet fertilization. Next, the farmer plans the fertilization of their field with the support of an FMIS with fertilization features (*FMIS 2*) that contains their *Field Geo-Data* and *Historic Yield Data* in the format required by it (*Field Geo-Data (Format 2)*) in its database. This task results in a *Fertilizer Application Map (Target)* that is also stored in *FMIS 2's* database (*Task: Plan Fertilization*). Subsequently, the farmer performs the fertilization manually based on the *Fertilizer Application Map (Target)*, which results in a *Fertilizer Application Map (Actual)* stored in *FMIS 2* (*Task: Perform Fertilization*).

Next, the farmer checks if the *Field Geo-Data* is up-to-date in *FMIS 3* (*Task: Check Field Geo-Data in FMIS 3*). If the *FMIS 3 Field Geo-Data* is out-of-date, the farmer then updates it in *FMIS 3*, which stores the *Field Geo-Data (Format 1) [Updated]* in its database. After this, they can move on to the planning of sugar beet plant protection (*Task: Update FMIS 3 Field Geo-Data*). If the *FMIS 3 Field Geo-Data* is up-to-date, the farmer can go directly to the planning of sugar beet plant protection. After that, the farmer plans the plant protection of their field with the assistance of an FMIS with plant protection features (*FMIS 3*) that contains their *Field Geo-Data* in the format required by it *(Field Geo-Data (Format 3))* in its database. This task results in a *Pesticide Application Map (Target)* that is stored in *FMIS 3's* database (*Task: Plan Plant Protection*). Then, the farmer performs the plant protection manually based on the *Pesticide Application Map (Target)*, which results in a *Pesticide Application Map (Actual)* that is stored in *FMIS 3* (*Task: Perform Plant Protection*).

Subsequently, the farmer plans the harvesting of their field manually (*Task: Plan Beet Harvesting*). Then, the farmer contracts a contractor (*Contractor 2*) for the harvesting work, who notifies the farmer when the job is done (*Task: Contract Harvesting from Contractor 2*). When the farmer receives the notification with the invoice for the completed work and the *Harvest Yield Map*, which is stored in *FMIS 2*, they pay *Contractor 2* for the work (*Task: Pay Contractor 2 for Beet Harvesting*).

Then, the farmer creates the nutrient flow balance with the aid of an FMIS (*FMIS 2*) with the *Fertilizer Application Map (Actual)* as input (*Task: Create Nutrient Flow Balance*). Next, the farmer sells the sugar beets to wholesalers based on the *Harvest Yield Map* (*Task: Sell Beets to Wholesalers*). The process ends when the beets are sold (*End Event: Beets Yield Sold*)."

## Process Analysis

With the as-is model ready, the process analysis step could start. The technique selected for this step was Waste Analysis, which aims at finding wasteful activities in a process from the perspective of the client. This technique relies on a qualitative investigation done by the process analyst, a role performed by the authors of this work.

When we inspected the as-is model, the activities that represent possible wastes were found to be *Check Field Geo-Data in FMIS * and the subsequent *Update FMIS * Field Geo-Data,* as they essentially deal with the same data, but in different systems. In reality, after having decided to plant sugar beets in a field, the farmer checks whether the currently stored field geographic data is the same as the one determined in

*Pacific Asia Conference on Information Systems 2022*

**11**

the field rotation planning beforehand. A field's geographic data, most importantly its boundaries, can be changed if the farmer decides to merge some fields into one or divide a field into smaller ones, depending on the farmer's needs. According to the domain expert consulted, this change is only done once before a new season. Therefore, it makes sense to perform this check only once and at the beginning of the process, before the seeding planning. Then the other tasks can be considered *overprocessing* waste of the data from the farmer's point of view.

Another issue we found in the process is that each FMIS used by the farmer has its own database and does not share any data with other FMISs, meaning the data is distributed. Furthermore, these FMISs demand some of the same data information about the farmer's field– in this case, the field geographic data– meaning that data is replicated and thus prone to future data inconsistencies affecting the process quality. To ensure that the process can return its best value, i.e., a good yield, the farmer needs to spend extra effort on keeping these FMISs consistent with the same updated data. This takes time and can potentially impact the process efficiency. Assuming all these tasks take the same amount of time x for checking and updating the system, regardless of the FMIS being checked or updated, this means that it can take the farmer up to three times (3x) as long to ensure data consistency throughout the process.

Furthermore, each of these FMISs uses a different format to represent the field's geographic data, as there is no set standard, which compromises the interoperability possibilities between them. These issues are documented in the issue register in Table 2.

| |
|---|
| **Issue 1:** Overprocessing of Field Geo-Data<br>**Description:** Farmers need to perform the same tasks up to three times during a process to ensure data consistency.<br>**Data and assumptions:** A farm has up to three independent FMISs to cover specific process activities. Checking and updating each at least once during a process represents up to three times the time actually needed for these tasks.<br>**Qualitative impact:** This overprocessing creates frustration for the farmer and can potentially affect the process efficiency.<br>**Quantitative impact:** Not applicable |
| **Issue 2:** Data Distribution<br>**Description:** Data is distributed across FMISs and other farm systems that do not share data with each other.<br>**Data and assumptions:** Farm systems have at least one kind of data in common, e.g., field geodata, each with its own copy.<br>**Qualitative impact:** This increases the probability of data inconsistencies between the systems throughout the process.<br>**Quantitative impact:** Not applicable |
| **Issue 3:** Data Interoperability<br>**Description:** There is no common data format standard between farm systems.<br>**Data and assumptions:** Common data formats are Shapefiles, GeoJSON, and GeoTIFF and the conversion between them must be done carefully.<br>**Qualitative impact:** This compromises interoperability between the systems and prevents the farmer from being able to manually share data easily from one system to another.<br>**Quantitative impact:** Not applicable |

**Table 3. Issue register of sugar beet planning as-is process**

This analysis is in accordance with the issues already brought to the farming industry's attention by the farmers. The agriculture domain expert pointed out that the farmers find themselves having to maintain two to three FMISs in order to have all their process activities covered. Due to the large size of the agricultural domain, having specialized FMISs is not the issue per se, but their lack of interest in sharing data with each other is (EIP-AGRI 2015). The system providers do not see any benefit for them, as it is a very costly development operation to change their systems and databases to be compatible with each other. Also, the farmer still pays for their service due to the lack of a better option.

# Process Redesign

This work used the Redesign Orbit to filter down which redesign methods would fit best to this case. Considering that the basis of this work is the redesign of the current sugar beet business process previously discovered and analyzed, redesign methods of a transactional nature are best suited for this case. Moreover, since the process is based on the farmer's perspective, the method should also be outward-looking. Finally, an analytical approach should be taken, as it is easier to verify and validate since it involves statistics of the process to analyze its characteristics (Dumas et al. 2018). Our analysis filtered the redesign methods down to three options: *Benchmarking*, *ERP-driven Redesign, and Lean.* In this work, we utilized the Lean method because it focuses on the added value of each activity, while other methods comprise mainly standard processes for enterprises (Dumas et al. 2018). The Lean method was popularized by Womack, Jones, and Roos (1990). In accordance with the Waste Analysis method used in this work, the Lean technique aims at waste elimination in a process by assessing whether an activity generates value from the perspective of a customer.

Along with the selected redesign methodology, there are several redesign heuristics that can be used to support the redesign process. Reijers and Mansar (2005) did an overview and qualitative evaluation of the most successful heuristics in order to define best practices in business process redesign. Based on the analysis results, the *Integral Technology* heuristic can be used to solve *Issue 2: Data Distribution*, since it relies mostly on the databases utilized by FMISs. This practice is technology-oriented and implies the application of new technology to elevate physical constraints in a process, thus achieving some kind of advantage, e.g., better quality of service or less time spent on tasks (Reijers and Mansar 2005). This would be done by integrating the digital field twin into the ADS Platform to be accessed by the different agricultural systems utilized by the farmer.

With a field twin, *Issue 3: Data Interoperability* would also be resolved as the ADS Platform is envisioned to enforce data standards or, in their absence, interpret the different data formats in order to provide better interoperability among the systems in the ecosystem. Furthermore, with the data being centralized in the field twin, *Issue 1: Field Geo-Data Overprocessing* is solved since the need to check the field geographic data more than once in a sugar beet process is eliminated. *Task elimination* is the redesign heuristic that makes the most sense in this case, as a task is considered unnecessary when it adds no value from the customer's point of view (Reijers and Mansar 2005). *Task redundancy* is defined as a special case of task elimination, which fits exactly to the sugar beet use case studied in this work. This heuristic focuses on business process operation, e.g., how the workflow is implemented, which seeks to improve the process speed and reduce its cost, sometimes with the consequence of reducing the process quality. However, in this case, the field twin ensures data consistency, which is the main concern identified for the quality of field geographic data in the sugar beet use case.

Based on the issues found by the analysis step and on the solutions researched in the project, the as-is process was redesigned into a to-be process. The model was syntactically verified by two BPM experts and the modeling tool. Moreover, semantic quality was validated by a COGNAC expert from Fraunhofer, who assured that the model fits the project's vision. Now, the field geographic data is checked only at the beginning of the process on the field twin, before seeding takes place, and all subsequent tasks are done without the need for data consistency checks. Therefore, for simplicity purposes, we only show the first paragraph of the to-be model's textual description that presents the field twin. The final version of the to-be business process model created is shown in Figure 4 below.

"The *Sugar Beet Farming To-Be* (Figure 4) process starts when the farmer decides to plant sugar beet in the field (*Start Event: Decision to Plant Beet Made*). Next, the farmer checks if the *Field Geo-Data* is up-to-date in the Field Twin, using an FMIS of their choice (*Task: Check Field Geo-Data*). The *Field Geo-Data* can either be up-to-date or out-of-date in the Field Twin according to the field geographic data desired for the current sugar beet farming. If the *Field Geo-Data* is out-of-date, the farmer then updates it in any FMIS of their choice, which stores it in the Field Twin. After that, they can move on to the planning of sugar beet seeding (*Task: Update Field Twin Geo-Data*). If the *Field Geo-Data* is up-to-date, the farmer can go directly to the planning of sugar beet seeding. [...]."

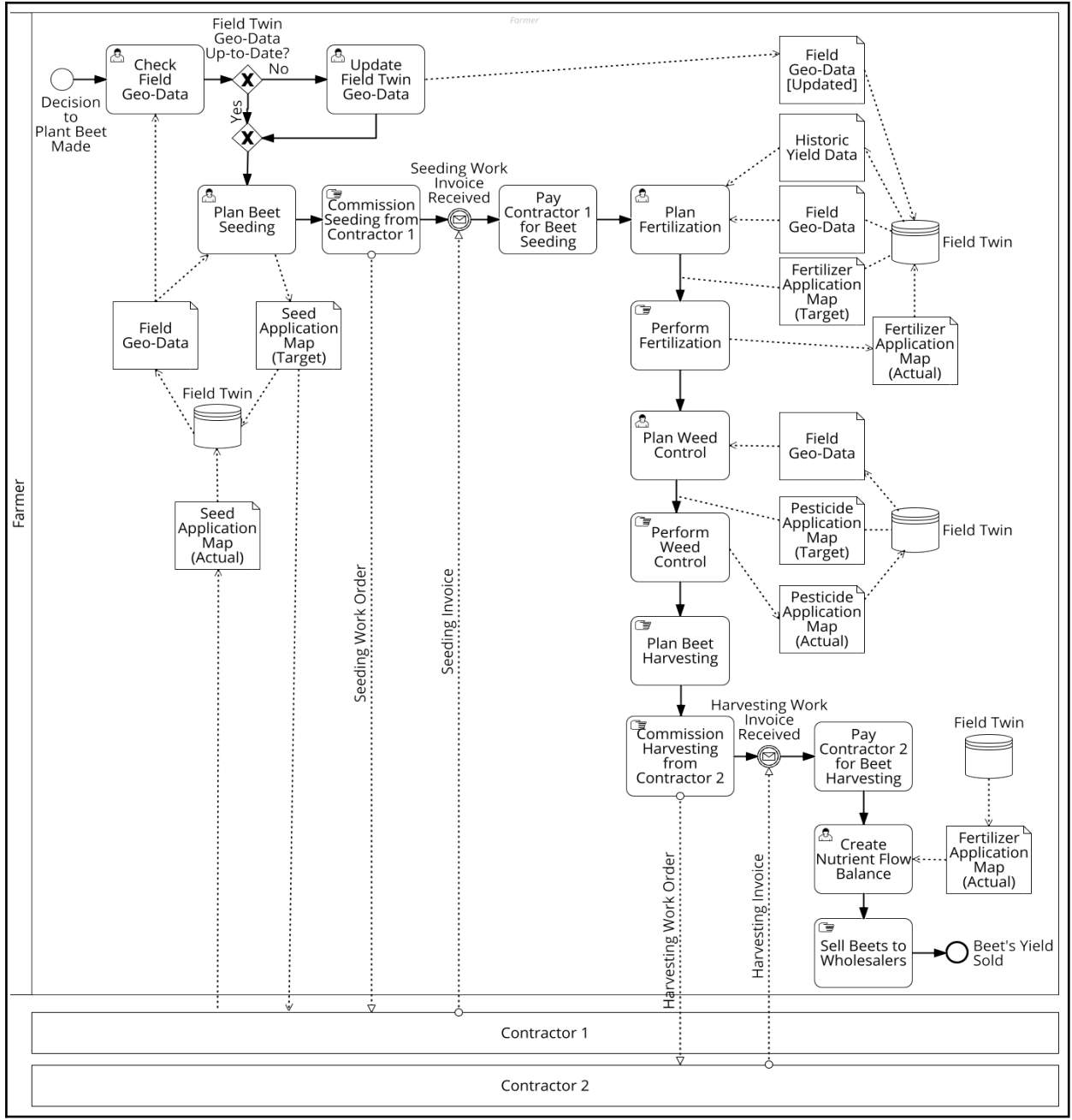

**Figure 4. Sugar beet farming to-be process model.**

# Evaluation and Discussion

For the comparison of the modeled processes, a new interview was conducted with the domain expert to estimate the processing time of the activities to be performed. The expert pointed out the complications in gathering time and cost data for each activity as this is something very intrinsic to each farm and depends on several variables such as employees, machines, and field sizes. Therefore, together with the expert, the times were estimated to reflect an average farm in Germany in the best possible way. The results are presented in Table 3, which shows how long each activity takes from start to finish without waiting times. These times were estimated assuming the farmer already knows how to use their FMISs and already has known contractors and wholesalers to get in contact with. Furthermore, data from KTBL, a widely used database for performance and cost accounting for crop production in Germany, was used to contribute to the estimation (KTBL 2022). The field size was assumed to be ten hectares, which is an average size field in Germany, according to the domain expert.

| Activity | Estimated Processing Time (min) |
|---|---:|
| Check Field Geo-Data in FMIS * | 05:00 |
| Update FMIS * Field Geo-Data | 15:00 |
| Plan Beet Seeding | 15:00 |
| Contract Service from Contractor | 07:00 |
| Pay Contractor for Service | 05:00 |
| Plan Fertilization | 25:00 |
| Perform Fertilization | 03:20:00 |
| Plan Plant Protection | 40:00 |
| Perform Plant Protection | 02:36:00 |
| Plan Beet Harvesting | 20:00 |
| Create Nutrient Flow Balance | 10:00 |
| Sell Beet to Wholesalers | 30:00 |

**Table 4: Estimated times for the sugar beet process activities**

To simulate the as-is and the to-be process with the estimated execution times obtained, this work utilized the business process tool Signavio. It provides customizable simulation scenarios and configuration of the duration time of each task. Also, it can receive the occurrence probability of exclusive gateway paths. In this work, the occurrence probability of the field geodata being out-of-date was five per cent, as mentioned by our domain expert, since farmers do not change their field boundaries very often. In both models, the best-case scenario is when the field geographic data is consistent with the planned state and no *Update FMIS Field Geo-Data* activity execution is necessary, while the worst-case scenario is when the field geographic data does not match with what was planned. In the as-is model, this means that all three FMISs are inconsistent and need to be updated accordingly. The average processing time is calculated by summing up all the activities' processing times and utilizing the weighted average at the decision points to update the field geographic data using the occurrence probability as the weight. Therefore, since the probability of needing to update the field data is five per cent, we sum up 0.05 × 15 for the average. The simulation results are shown in Table 4.

| Model | Best-Case Scenario | Worst-Case Scenario | Average Processing Time |
|---|---|---|---:|
| as-is | 08:55:00 | 09:40:00 | 08:57:15 |
| to-be | 08:45:00 | 09:00:00 | 08:45:45 |

**Table 5. Process simulation results.**

As shown in Table 4, the to-be process presents a processing time improvement in all cases compared to the as-is process, especially when considering the worst-case scenario. This not only represents an efficiency improvement by reducing the overprocessing of the field data, but also a data quality improvement of the process since the field data is centralized in the field twin. This means that the process is much less prone to data inconsistencies and interoperability problems.

As observed throughout this work, the issues and solutions found for the sugar beet use case revolve around qualitative aspects of the process, namely, interoperability and data quality. The lack of interoperability between the systems used in a farm process leads to data being distributed and creates the need for extra activities without real value for the farmer to ensure the data is consistent during the whole season. This is detrimental to the overall process efficiency. Moreover, it prevents the farmer and the systems themselves from leveraging all the gathered and processed data and improving the processes and their services as a whole. The ADS proposed by our project, which envisions the use of digital field twins, tackles these issues by centralizing the data, making it semantically interoperable by enforcing format standards and interpreting other formats when possible. Furthermore, it enforces collaboration between existing systems to use and update the field twin data, allowing the farmer to authorize their systems to access the DT data, which would always be consistent as long as the connection to the DTs is kept.

## Conclusion

The proposal of COGNAC to integrate DTs into the agricultural domain through the ADS Platform in order to achieve greater interoperability among systems and ecosystems was demonstrated and evaluated in the context of a sugar beet farming use case. The use case was discovered according to the common way this process is done currently in Germany. Three steps from the BPM lifecycle—Process Discovery, Process Analysis, and Process Redesign—were utilized to model and redesign it. The current process was discovered and semantically validated using a semi-structured interview method with the project's domain expert, generating the as-is process model in BPMN along with the textual process description. Subsequently, the as-is model was analyzed following the Waste Analysis methodology, resulting in an issue register table to be used in the redesign step. Based on the issue register and the as-is model, the process was redesigned according to the Lean methodology and COGNAC's vision, generating the to-be process model including the ADS Platform and the DT concept. Finally, both the as-is and the to-be process were simulated based on their estimated activity processing time to compare their performances.

The study using the BPM lifecycle steps methodology emphasizes the need for more interoperability and data quality in the current ADS, which includes systems, services, and digital ecosystems used by farmers. This was observed with the process analysis step, where issues already pointed out by farmers were identified and thereby confirmed. With respect to our RQ, we found in the process obtained from the redesign step that the usage of the DT improved the interoperability among the systems in the sugar beet process. It was achieved through the enforcement of data format standards and the centralization of data in the digital field twin. Additionally, the enhanced process shows a reduction in the number of tasks needed to be performed by the farmer, which leads to an efficiency improvement as demonstrated by the simulation of both models in the evaluation phase. These results corroborate the potential of adopting DTs for the interoperability improvement in the ADS. With more interoperability among its systems and services, farmers and agriculture in general would benefit from being able to use their entire data for e.g. simulations, forecasting, and analysis purposes.

Future work will comprise applying the BPM steps to other use cases relevant to the agricultural domain so that they can be further analyzed and improved accordingly. This will serve to evaluate whether the concepts proposed in the project also meet the needs of other use cases. Moreover, the implementation of the ADS Platform and the digital field twins will be done to further validate this concept.

## Acknowledgements

This study was financed by the Coordenação de Aperfeiçoamento de Pessoal de Nível Superior - Brasil (CAPES) - Finance Code 001. We also thank Dr. Jan Mendling for his insights into our work.

## References

Avila, D. T., Thom, L. H. 2021. "Process Modeling Guidelines". In *Business Process Management Journal*, v. 27, p. 1-23 (doi: 10.1108/BPMJ-10-2019-0407).

Alves, R. G. et al. 2019. "A digital twin for smart farming". *IEEE Global Humanitarian Technology Conference*, GHTC 2019. (doi: 10.1109/GHTC46095.2019.9033075)

Celestrini, J. R. et al. 2019. "An architecture and its tools for integrating iot and bpmn in agriculture scenarios". In *Association for Computing Machinery*, Part F147772, p. 824–831. (ISBN: 9781450359337).

Cestari, R. H., Ducos, S., & Exposito, E. 2020. "IPaaS in Agriculture 4.0: An Industrial Case". In *Proceedings of the Workshop on Enabling Technologies: Infrastructure for Collaborative Enterprises*, WETICE, 2020-September, 48–53. (doi: 10.1109/WETICE49692.2020.00018)

Dias, C. L. de B. et al. 2019. "Anti-patterns for process modeling problems: An analysis of bpmn 2.0-based tools behavior". *Springer*, v. 362 LNBIP, p. 745–757. (ISBN: 9783030374525). (ISSN: 18651356).

Dörr, J. and Nachtmann, M. 2022. "Handbook Digital Farming" (1st ed.). *Springer*. (ISBN: 978-3-662-64377-8).

Dumas, M., Rosa, M. L., Mendling, J., & Reijers, H. A. 2018. "Fundamentals of Business Process Management (2nd ed.)". *Springer*. (doi: 10.1007/978-3-662-56509-4).

EIP-AGRI. 2015. "Eip-agri focus group precision farming final report". (Accessed: 03/10/21). (https://ec.europa.eu/eip/agriculture/sites/default/files/eip-agri_focus_group_on_precision_farming_final_report_2015.pdf).

Grieves, Michael. 2014. "Digital Twin: Manufacturing Excellence through Virtual Factory Replication" *White paper* 1(2014) : 1-7. (https://www.researchgate.net/publication/275211047)

IESE, F. 2020. "Agricultural Data Space (Ads) A Publication Of The Fraunhofer Lighthouse Project "Cognitive Agriculture"". (https://www.iese.fraunhofer.de/content/dam/iese/de/dokumente/innovationsthemen/COGNAC_Whitepaper_ADS_eng.pdf).

IESE, F. 2021. "Digital Twins for Agriculture". (Accessed: 27/09/21). (https://www.iese.fraunhofer.de/blog/digital-twins-agriculture/).

Keates, O. 2019. "Integrating iot with bpm to provide value to cattle farmers in Australia". In *Remco*; Chiara, Z. U. D. F.; Dijkman (Ed.). Springer International Publishing, p. 119–129. (ISBN: 978-3-030-37453-2).

KTBL. 2022. "About us". (Accessed: 26/02/2022). (https://www.ktbl.de/wir/wir).

Mendling, J., Reijers, H., & Aalst, W. 2010. "Seven Process Modeling Guidelines (7PMG)", in *Information and Software Technology 52,* issue 2, February, Butterworth-Heinemann, pp. 127-136. (doi: 10.1016/j.infsof.2009.08.004).

OMG. 2021. "Business Process Model And Notation". (Accessed: 02/10/21). (https://www.omg.org/spec/BPMN/2.0/About-BPMN/).

OMG. 2022. "Graphical Notations for Business Processes". (Accessed: 12/05/22). (https://www.omg.org/bpmn/).

Pylianidis, C., Osinga, S., & Athanasiadis, I. N. 2021. "Introducing digital twins to agriculture". *Computers and Electronics in Agriculture*, 184. (doi: 10.1016/j.compag.2020.105942)

Qi, Q. et al. 2021. "Enabling technologies and tools for digital twin". *Journal of Manufacturing Systems,* Elsevier, v. 58, p. 3–21, 1 (ISSN: 0278-6125).

Rasheed, A., San, O., Kvamsdal, T. 2020. "Digital twin: Values, challenges and enablers from a modeling perspective". *IEEE Access, Institute of Electrical and Electronics Engineers Inc.*, v. 8, p. 21980–22012, (ISSN: 21693536).

Reijers, H. A.; Mansar, S. L. 2005. "Best practices in business process redesign: An overview and qualitative evaluation of successful redesign heuristics". *Omega*, Elsevier BV, v. 33, p. 283–306, (ISSN: 03050483).

Rupnik, R., Vavpotič, D., Jaklič, J., Kuhar, A., Plavši´c, M. P., Žvanut, B., & Rodrigues, G. C. 2021. "A Reference Standard Process Model for Agriculture to Facilitate Efficient Implementation and Adoption of Precision Agriculture". (https://doi.org/10.3390/agriculture)

Soares Silva, T., Toralles Avila, D., Ampos Flesch, J., Marques Peres, S., Mendling, J., & Thom, L. H. 2019. "A Service-Oriented Architecture for Generating Sound Process Descriptions", in 2019 IEEE 23rd *International Enterprise Distributed Object Computing Conference*, Paris, France, pp. 1-10. (doi: 10.1109/EDOC.2019.00011).

Tao, F. et al. 2019. "Digital twin in industry: State-of-the-art". *IEEE Transactions on Industrial Informatics*, v. 15, 4 (ISSN: 1551-3203).

Haße, H., van der Valk, H., Weissenberg, N., Otto, B. 2020. "Shared Digital Twins: Data Sovereignty in Logistics Networks Shared Digital Twins: Data Sovereignty in Lo-gistics Networks". (https://www.researchgate.net/publication/344437918)

Womack, J. P., Jones, D. T., Roos, D. 1990. "The machine that changed the world: the story of lean production – Toyota's secret weapon in the global car wars that is revolutionizing world industry".

Zaninelli, M., Pace, M. R. 2018. "The o3-farm project: First evaluation of a business process management (bpm) approach through the development of an experimental farm management system for milk traceability". Agriculture (Switzerland), MDPI AG, v. 8, 9 (ISSN: 20770472).